\documentclass[prl,reprint,groupedaddress]{revtex4-1}
\usepackage{graphicx}
\usepackage{dcolumn}
\usepackage{amsmath}
\usepackage{amsfonts}
\usepackage{amssymb}
\usepackage{epsfig,float,afterpage,wrapfig,psfrag}
\usepackage{natbib}

\newcommand{\beq}{\begin{equation}}
\newcommand{\eeq}{\end{equation}}
\newcommand{\bes}{\begin{subequations}}
\newcommand{\ees}{\end{subequations}}
\newcommand{\bea}{\begin{eqnarray}}
\newcommand{\eea}{\end{eqnarray}}
\newcommand{\ba}{\begin{array}}
\newcommand{\ea}{\end{array}}
\newcommand{\beqn}{\begin{eqnarray*}}
\newcommand{\eeqn}{\end{eqnarray*}}

\newcommand{\f}[2]{\frac{#1}{#2}}

\newcommand{\om}{\omega}
\newcommand{\G}{\Gamma}

\def\nn{\nonumber}

\newlength{\sizeonefig}
\newlength{\sizetwofig}
\begin{document}
\title{Majorana fermions in an out-of-equilibrium topological superconducting wire: an exact microscopic transport analysis of a \textit{p}-wave open chain coupled to normal leads}
\author{Dibyendu Roy}
\author{C. J. Bolech}
\author{Nayana Shah}
\affiliation{Department of Physics, University of Cincinnati, OH 45221-0011}

\begin{abstract}
Topological superconductors are prime candidates for the implementation of topological-quantum-computation ideas because they can support non-Abelian excitations like Majorana fermions. We go beyond the low-energy effective-model descriptions of Majorana bound states (MBSs), to derive non-equilibrium transport properties of wire geometries of these systems in the presence of arbitrarily large applied voltages. Our approach involves quantum Langevin equations and non-equilibrium Green's functions. By virtue of a full microscopic calculation we are able to model the tunnel coupling between the superconducting wire and the metallic leads realistically; study the role of high-energy non-topological excitations; predict how the behavior compares for increasing number of odd \textit{vs.}~even number of sites; and study the evolution across the topological quantum phase transition (QPT). We find that the normalized spectral weight in the MBSs can be remarkably large and goes to zero continuously at the topological QPT. Our results have concrete implications for the experimental search and study of MBSs.
\end{abstract}

\maketitle

Topological quantum computing \cite{Nayak08} is interesting not only because of the technological goal but also because it requires the realization and manipulation of new types of quasiparticle excitations that have never been observed before and the question of their actual existence tests the limits of our understanding of the quantum world at a fundamental level. The simplest proposals along this line involve the realization of Majorana fermions (MFs).

One of the pressing theoretical problems in this context is to sort out the experimental signatures of MFs. Several proposals for their realization in (effective) one-dimensional (1D) topological (\textit{p}-wave) superconductors have been put forward recently \cite{Lutchyn10,*Oreg10,*Nersesyan11,*Sau11}. 
A natural way to probe superconducting (SC) nanowires is via transport measurements and STM spectroscopy. Such transport studies of mid-gap features in the spectrum of unconventional superconductors have consistently attracted considerable attention \cite{Kashiwaya00} and particularly so in the quasi-one-dimensional case of (possibly \textit{p}-wave) organic superconductors \cite{Sengupta01,Ha03}. The renewed interest in MBSs has given rise to more detailed studies including some exciting recent experimental claims of MBS detection \cite{Sasaki11}. 

Important insights have been gained from past theoretical works
\cite{Bolech07,Nilsson08,Flensberg10} that study low-energy effective models
consisting of the MBSs and an effective coupling to and between them as
proposed in Ref.~\onlinecite{Bolech07}. 
These works have made specific predictions for detecting MBSs in experiments. 
For instance, non-equilibrium electrical transport between a metallic probe and a collection of coupled MBSs has been derived \cite{Bolech07,Nilsson08,Flensberg10} and the differential electrical conductance calculated in the limit where the applied voltage ($eV$)
is much smaller than the SC energy gap.
It is crucial to go beyond low-energy approximations and calculate the full current-voltage characteristics including the contribution of the  non-topological higher-energy excitations.
It is also important to check the regime of validity of these low-energy models and even better have a full microscopic calculation to predict and interpret experimental results in realistic systems.

While different proposals for realizing MBSs in 1D differ in details  \cite{Lutchyn10,*Oreg10,*Nersesyan11,*Sau11}, all essential
features of realistic systems are expected to be minimally described by the so
called Kitaev chain (linked to the early theoretical proposals for topological
quantum computation) \cite{Kitaev00}. We thus study non-equilibrium electrical transport in Kitaev's tight-binding model of a
single-channel 1D quantum wire with triplet 
(\textit{p}-wave) SC pairing between electrons with the same spin orientation (equal-spin pairing). The model's spectrum includes two MBSs at the two ends of the wire, which are expected to exhibit non-Abelian braiding statistics (in wire networks \cite{Alicea11,*Clarke11,*Halperin11}), are separated from the non-topological complex excitations by an energy gap and have been proposed as a basis for a decoherence-free topological quantum memory. 
In view of theoretically capturing the physics of realistic experiments, we are
interested in studying non-equilibrium transport through the 1D quantum wire
in an open-quantum-system geometry, \textit{i.e.}, the wire is connected to
external baths which incorporate both decoherence and dissipation. This is a minimal setup that can capture the essential features of recent proposals for realizing \textit{p}-wave SC wires \cite{Lutchyn10,*Oreg10}. We develop a novel extension of the quantum Langevin Equations and Green's Function (LEGF) approach \cite{Dhar03} to SC wires with BCS-type pairing. LEGF provides us with an appropriate exact method to determine the electrical current in the SC wire with arbitrary values of $eV$ as compared with the
strengths of the contacts and the SC pairing amplitude. Although this method has proven quite useful to investigate electrical
\cite{Dhar03,DharSen06,Roy07} as well as thermal \cite{DharRoy06}
non-equilibrium transport in non-interacting, ordered and disordered quantum
systems connected to baths, in this letter we will focus only on the electrical transport.

The Hamiltonian of the full system is given by $\mathcal{H}=\mathcal{H}_\mathrm{W}+\mathcal{H}_\mathrm{B}^{L}
+\mathcal{H}_\mathrm{B}^{R}+\mathcal{H}_\mathrm{WB}^{L}
+\mathcal{H}_\mathrm{WB}^{R}$ where
\begin{align}
\mathcal{H}_\mathrm{W}  & =-\nu\sum_{l=1}^{N}(a_{l}^{\dag}a_{l}-\frac{1}{2}%
)+\sum_{l=1}^{N-1}\big[-\gamma~(a_{l}^{\dag}a_{l+1}+a_{l+1}^{\dag}%
a_{l})\nonumber\\
& +(|\Delta|e^{i\theta}a_{l}a_{l+1}+|\Delta|e^{-i\theta}a_{l+1}^{\dag}%
a_{l}^{\dag})\big]\label{Ham}\\
\mathcal{H}_\mathrm{B}^{p}  & =-\gamma_{p}~\sum_{\alpha=1}^{\infty}~(a_{\alpha
}^{p\dag}a_{\alpha+1}^{p}+a_{\alpha+1}^{p\dag}a_{\alpha}^{p}%
)~,~~~p=L,R\nonumber\\
\mathcal{H}_\mathrm{WB}^{p}  & =-\gamma_{p}^{\prime}~(a_{1}^{p\dag}a_{l_p}+a_{l_p}^{\dag
}a_{1}^{p})~,~~~l_L=1~,~l_R=N\nonumber
\end{align}
The term $\mathcal{H}_\mathrm{W}$ denotes the Hamiltonian of the 1D wire and $\mathcal{H}_\mathrm{B}^{p}$ that of the $p^{\mathrm{th}}$ bath, while the coupling between them is
$\mathcal{H}_\mathrm{WB}^{p}$. Here $a_{l}$ and $a_{\alpha}^{p}$ denote standard fermion operators on the SC wire and on the $p^{\rm th}$ bath, respectively. The \textit{p}-wave BCS-pairing amplitude of the wire is $\Delta$ and the parameter $\nu$ can be tuned to drive the SC wire in or out of a topological phase (cf.~Refs.~\onlinecite{Kitaev00,Motrunich01}).

\begin{figure}[htb]
\begin{center}
\includegraphics[width=8.0cm]{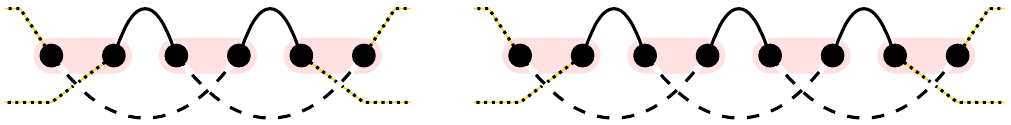}
\end{center}
\caption{Schematic depiction of a Kitaev chain in terms of MFs (filled circles) for $\nu=0$. Two generic cases with odd ($N=3$) and even ($N=4$) number of sites are shown side by side. The solid and dashed lines correspond to $|\Delta|\pm\gamma$ matrix elements, respectively. The two dotted lines at each end represent equal-strength contacts with the respective lead. Since for $\nu=0$ there is no matrix element between the two MFs within each site, one can see the existence of two independent Majorana (conduction) channels with $N$ MFs each. Notice those two channels are always mirror-equivalent for odd $N$ and contain an exact zero mode each. For $|\Delta|=\gamma$ and any odd or even $N$, the dashed lines are absent and there are two zero-energy MBSs localized one at each end and decoupled from the rest of the chain.}
\label{MajChain} 
\end{figure}

Next we introduce a local re-parametrization for the electron
operators on the SC wire in terms of MFs \cite{Kitaev00}, $a_{l}=e^{-i\theta/2}(c_{2l-1}+i\,c_{2l})/2$, with
$c_{m}^{\dag}=c_{m}$ and $\{c_{m},c_{m^{\prime}}\}=2\delta_{m,m^{\prime}}\,$.
The wire Hamiltonian in the MF basis (see Fig.~\ref{MajChain}) is given by $\mathcal{H}_\mathrm{W}
=(i/2)\big[(-\nu)\sum_{l=1}^{N}c_{2l-1}c_{2l}
+\sum_{l=1}^{N-1}\big\{(|\Delta|+\gamma)~c_{2l}c_{2l+1}
+(|\Delta|-\gamma)c_{2l-1}c_{2l+2}\big\}\big]$. 
The two MFs at any one end of the wire are connected to the end site of the bath with equal-strength coupling. 
(Note that the contact coupling to the inner MF of the end site of the wire is outside the scope of all previous works, as they deal either with effective MBS models \cite{Bolech07,Nilsson08,Flensberg10} or a continuum representation \cite{Sengupta01}.) 
We assume that for times $t\leq t_{0}$, the wire is disconnected from the
baths (each in equilibrium at a specified temperature $T_{p}$ and
chemical potential $\mu_{p}$). At time $t_{0}$ we connect both baths to the 1D SC wire and we are interested in the steady-state properties,
which can be derived by integrating out the bath
operators using the quantum LEGF formalism and considering the limit
$t_{0}\rightarrow-\infty$. We get the following solution
\begin{align}
& \tilde{c}_{l}(\omega)=\sum_{m=1}^{2N}~G_{lm}^{+}(\omega)~\tilde{h}%
_{m}(\omega)~~~\mathrm{for}~~l=1,2,..2N,\label{sol}\\
& G^{+}(\omega)=Z^{-1}(\omega)~~\mathrm{and}~~Z_{lm}(\omega)=\Phi_{lm}%
(\omega)+A_{lm}(\omega),\nonumber\\
& \Phi_{lm}(\omega)=\omega~\delta_{l,m}+\frac{i\nu}{\hbar}\delta_{l,m-1}%
\delta_{l,\mathrm{odd}}-\frac{i\nu}{\hbar}\delta_{l,m+1}\delta
_{l,\mathrm{even}}\nonumber\\
&\qquad\qquad +\frac{i(|\Delta|+\gamma)}{\hbar}
(\delta_{l,m+1}\delta_{l,\mathrm{odd}}
-\delta_{l,m-1}\delta_{l,\mathrm{even}})\nonumber\\
&\qquad\qquad -\frac{i(|\Delta|-\gamma)}{\hbar}
(\delta_{l,m-3}\delta_{l,\mathrm{odd}}
-\delta_{l,m+3}\delta_{l,\mathrm{even}})~,\nonumber\\
& A_{lm}(\omega)=-\delta_{l,m}[(\delta_{l,1}+\delta_{l,2})\Sigma_{L}^{+}
+(\delta_{l,2N-1}+\delta_{l,2N})\Sigma_{R}^{+}]~,\nonumber\\
& \tilde{h}_{m}(\omega)=\big[\tilde{\eta}_{L}(\omega)-\tilde{\eta}_{L}^{\dag
}(-\omega)]\delta_{m,1}-i\big[\tilde{\eta}_{L}(\omega)+\tilde{\eta}_{L}^{\dag
}(-\omega)]\delta_{m,2}\nonumber\\
& +\big[\tilde{\eta}_{R}(\omega)-\tilde{\eta}_{R}^{\dag}(-\omega
)]\delta_{m,2N-1}-i\big[\tilde{\eta}_{R}(\omega)+\tilde{\eta}_{R}^{\dag
}(-\omega)]\delta_{m,2N}~,\nonumber
\end{align}
where $\tilde{c}_{m}(\omega)=(1/2\pi)\int_{-\infty}^{\infty}dt~
e^{i\omega t}c_{m}(t)$ and
same for $\tilde{\eta}_{p}(\omega)$.
The single-particle retarded matrix Green's function of the $p^{\mathrm{th}}$ bath is given by $g^{p+}(t) =-i\theta(t) e^{-i\mathcal{H}_\mathrm{B}^{p}t/\hbar}$ and $g_{ll'}^{p+} (\omega)=\int_{-\infty}^{\infty}dt~e^{i\omega t}g_{ll'}^{p+}(t)$.
Its first diagonal element  (in the MF `site' basis) is used to define $\Sigma_{p}^{+}(\omega)=(\gamma_{p}^{\prime}/\hbar)^{2}g_{11}^{p+}(\omega)$,
which corresponds to dissipation due to coupling with the bath. The `noise' in the leads,
\bea
\eta_{p}(t)=-\frac{i\gamma_{p}^{\prime}e^{i\theta/2}}{\hbar}~\sum_{\alpha
=1}^{\infty}g_{1\alpha}^{p+}(t-t_{0})~a_{\alpha}^{p}(t_{0})\label{noise}
\eea
depends on the bath's initial distribution and obeys \cite{DharSen06}
\bea
\langle\tilde{\eta}_{p}^{\dag}(\omega)\tilde{\eta}_{m}(\omega^{\prime}%
)\rangle=\Gamma_{p}(\omega)~f(\omega,~\mu_{p},~T_{p})~\delta(\omega
-\omega^{\prime})~\delta_{pm}\;\label{nncor}
\eea
where $f(\omega,\mu,T)$
is the Fermi distribution function. The above identity is a form of fluctuation-dissipation relation. We use the definition $\Gamma_{p}=-\mathrm{Im}[\Sigma_{p}^{+}]/\pi=(\gamma_{p}^{\prime}/\hbar)^{2}\rho_{p}$ where $\rho_{p}(\omega)$ is the local density of states at the first site on the
$p^{\mathrm{th}}$ bath. Finally, $G^{+}(\omega)$ is the Green's function of
the full system (wire plus baths) \cite{DharSen06}.

The outward current between the wire and the $p^\mathrm{th}$ baths is given by
\bea
j_{p}(t)=-\f{i\gamma_{p}^{\prime}}{\hbar}\langle~a_{l_p}^{\dag}a_{1}^{p}
-{a_{1}^{p}}^{\dag}a_{l_p}~\rangle
=2~\mathrm{Im}\big[\f{\gamma_{p}^{\prime}}{\hbar}\langle~a_{l_p}^{\dag}a_{1}^{p}\rangle\big]
\eea
where $\langle...\rangle$ denotes averaging over noise like
in Eq.~(\ref{nncor}).
For the sake of concreteness and simplicity of presentation, we will consider the `symmetric' case for which the two baths to are similar ($\gamma_{L}=\gamma_{R}$) and connected with the SC wire by identical contacts ($\gamma^{\prime}_{L}=\gamma^{\prime
}_{R}$ or $\Gamma_{L}=\Gamma_{R}\equiv\Gamma$) while $\mu_{L}=\tilde{\mu}/2=-\mu_{R}$ and $T_{L}=T=T_{R}$. 
In this case the steady-state currents $j_{\mathrm L,R}$ are equal up to a sign and
\begin{widetext}
\begin{eqnarray}
&&j_{\mathrm L}(\tilde{\mu}/2,-\tilde{\mu}/2)=\int_{-\infty}^{\infty}d\omega~\Big(\sum_{l,k,m,n=1,2}(-1)^{|l-k|}i^{(l+k+m+n)}G^+_{lk}(-\om)G^+_{mn}(\om)+\sum_{l,m=1,2}^{k,n=2N-1,2N}(-1)^{(k-l)}(-i)^{(k+n-l-m)}\nn \\&& G^+_{lk}(-\om)G^+_{mn}(\om)\Big)\big(\f{-\pi}{2}\big)\G^2(\om)(f(\om,\tilde{\mu}/2,T)-f(\om,-\tilde{\mu}/2,T))\equiv\int_{-\infty}^{\infty}\f{d\omega}{2\pi} \mathcal{T}(\om)(f(\om,\tilde{\mu}/2,T)-f(\om,-\tilde{\mu}/2,T)) \label{lsCurr}
\end{eqnarray}
\end{widetext}
We have checked that the expression in Eq.~(\ref{lsCurr}) reduces to the known result for the case of a normal metallic wire ($\Delta=0$) where $\mathcal{T}(\omega)$ is the transmission coefficient of a single electron through the wire.
For a SC wire, $\mathcal{T}(\omega)$ that contains contributions to transport coming from both electrons and holes, can be simply interpreted in the symmetric case via the zero-temperature differential conductance (with $\tilde{\mu}\equiv eV$, $e=1$)
\begin{equation}
\frac{dI}{dV}=\frac{dj_{\mathrm L}(\tilde{\mu}/2,-\tilde{\mu}/2)}{d\tilde{\mu}%
}=\frac{1}{4\pi}\Big(\mathcal{T}(\frac{\tilde{\mu}}{2})+\mathcal{T}(\frac
{-\tilde{\mu}}{2})\Big)
\end{equation}
Moreover, for $\nu=0$, $\mathcal{T}(\omega)$ is an even function and thus proportional to $dI/dV$.

\begin{figure}[htb]
\begin{center}
\includegraphics[width=8.0cm]{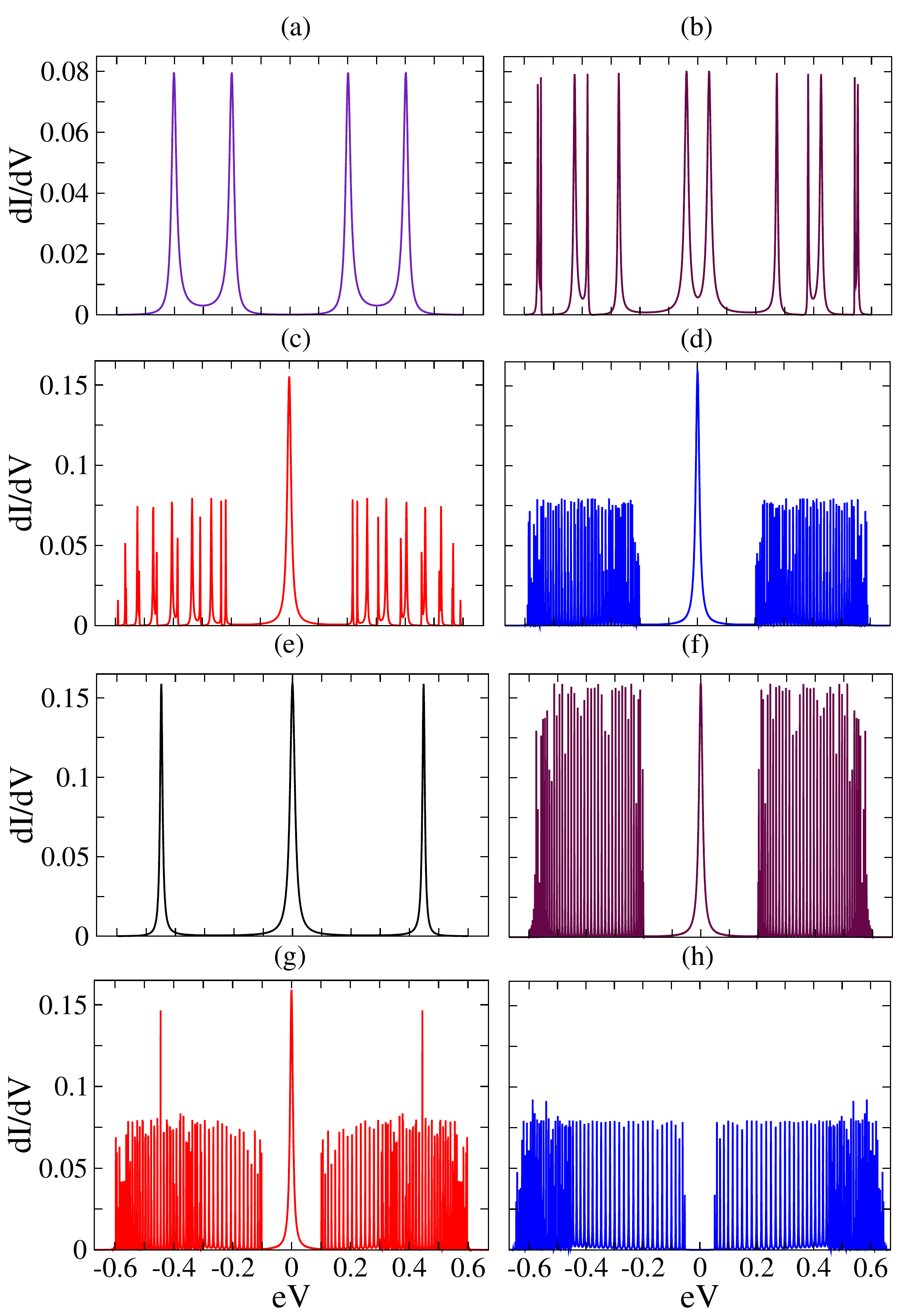}
\end{center}
\caption{$dI/dV$ (in units of $e^2/\hbar$) \textit{vs.} $eV$ for $\gamma_{p}=1$, $\gamma'_{p}=\gamma=0.1$ ($p=L,R$) and $|\Delta|=0.3$. Everywhere $\nu=0$ except for $\nu=0.1, 0.25$ in (g,h) respectively. Wire lengths are as follows: (a) $N=2$, (b) $N=6$, (c) $N=16$, (d) $N=100$, and (e) $N=3$, (f-h) $N=101$.}
\label{dc1} 
\end{figure}

At $\nu=0$ and $\gamma=\Delta$, there exist two exactly zero-energy localized MBSs (decoupled from rest of the wire) and the corresponding zero-temperature $dI/dV$ has a peak at zero bias and two more peaks in $dI/dV$ at $eV=\pm(\gamma+\Delta)$ coming from all the degenerate high-energy excitations at these parameter sets. As can be seen from Fig.~\ref{dc1}(a-f), the $dI/dV$ shows distinct behaviors for finite $N$ depending on whether $N$ is odd (panels e,f) or even (panels a-d) once $\gamma\ne\Delta$. We have developed a physical understanding of different behaviors by closely looking at the behavior of a finite Kitaev chain with odd \textit{vs.}~even $N$ (see Fig.~\ref{MajChain}). For odd $N$, there are $N$ pairs of degenerate modes including a pair of zero-energy modes thereby showing $N$ peaks (including one at $V=0$). 
For even $N$, all modes, including the ones at zero energy are split and the zero-temperature $dI/dV$ has a dip at zero-bias. However, the splitting in the MBSs decreases with increasing wire length, and
finally goes to zero in the long-wire limit as $\exp(-\alpha N)$ with $\alpha\sim(\Delta+\gamma)/|\Delta-\gamma|$ for $\Delta\sim\gamma$. 
(This is in contrast to the result of an effective model of a chain of coupled MBSs, for which the  zero-bias conductance is zero for the case of two MBSs \cite{Flensberg10}.)
Note that the splitting of the non-topological high-energy excitations survives for a long even-site wire as
can be seen by comparing heights in Fig.~\ref{dc1}(a-d).

For a finite $\nu$, the high-energy excitations are split for any even or odd number of sites \footnote{We have checked that the isolated spikes are not numerical artifacts and that they can be traced back to the structure of $\mathcal{T}(\omega)$.}. Gating the wire ($0<|\nu|<2|\gamma|$) also splits any degenerate zero modes at $V=0$ for the finite $N$ but, remarkably, in the large $N$ limit, they regain their integrity and yield a zero-bias peak. (The splitting for finite-$N$ case can be understood by adding a matrix element between the two MFs within each site in Fig.~\ref{MajChain}.) Next, in the long-wire case, we closely study how $dI/dV$ evolves as $\nu$ is varied across the topological QPT [\textit{i.e.}~the quantum phase transition from the topological ($|\nu|<2|\gamma|$) to non-topological ($|\nu|>2|\gamma|$) SC phase]. 
We observe that as $|\nu|$ approaches $2|\gamma|$, the gap shrinks and the zero-bias peak corresponding to the MBSs loses weight; at $|\nu|=2|\gamma|$ the gap closes completely; and for $|\nu|>2|\gamma|$ the gap reopens with no zero-bias anomaly [see Fig.~\ref{dc1}(h)].

\begin{figure}[htb]
\begin{center}
\includegraphics[width=8.0cm]{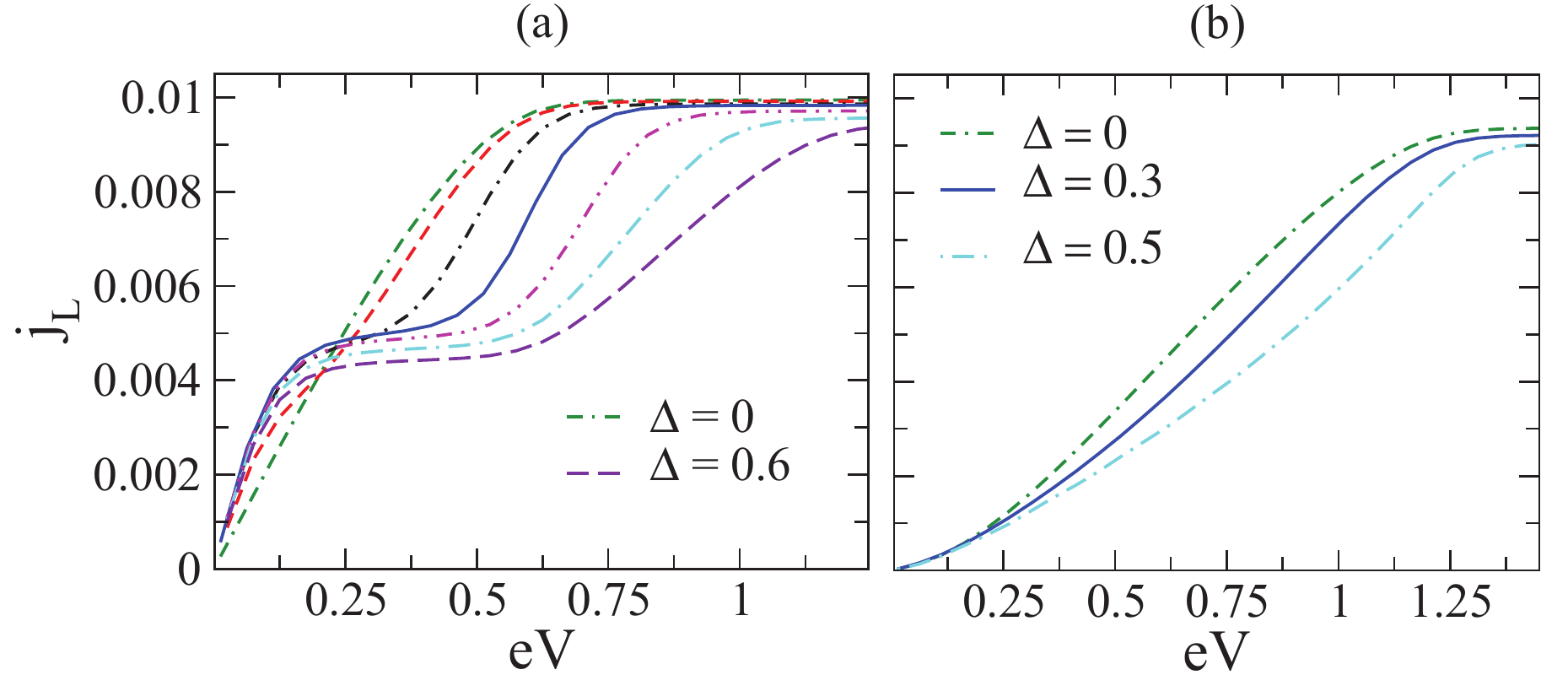}
\end{center}
\caption{I-V characteristics for a symmetrically biased wire. $N=15$, $\gamma_{p}=10\,\gamma'_{p}=1$ ($p=L,R$) and $\gamma=0.3$. In (a) $\Delta$ changes as $0, 0.1, ... , 0.6$ while $\nu=0$; and in (b) $\nu=0.65$.}
\label{I-V}
\end{figure}

In Fig.~\ref{I-V}(a) we compare the I-V characteristics of the wire for different parameters (varying $\Delta$).
In the topological phase, $I$ displays a sharp rise for small voltages due to the presence of the low-energy MBSs and then saturates with increasing voltage reflecting the gap in the energy-spectrum. The gap and hence the plateau is most sharply defined for $|\Delta|\gtrsim|\gamma|$.
The current increases again for even higher voltages, when non-topological excitations in the SC wire start to contribute to the transport, and finally saturates again when the applied voltage is larger than the band-width of the wire.
For comparison, we plot in Fig.~\ref{I-V}(b) the I-V of the topologically trivial SC phase along with the corresponding metallic phase with the same $\nu$.

\begin{figure}[htb]
\begin{center}
\includegraphics[width=8.0cm]{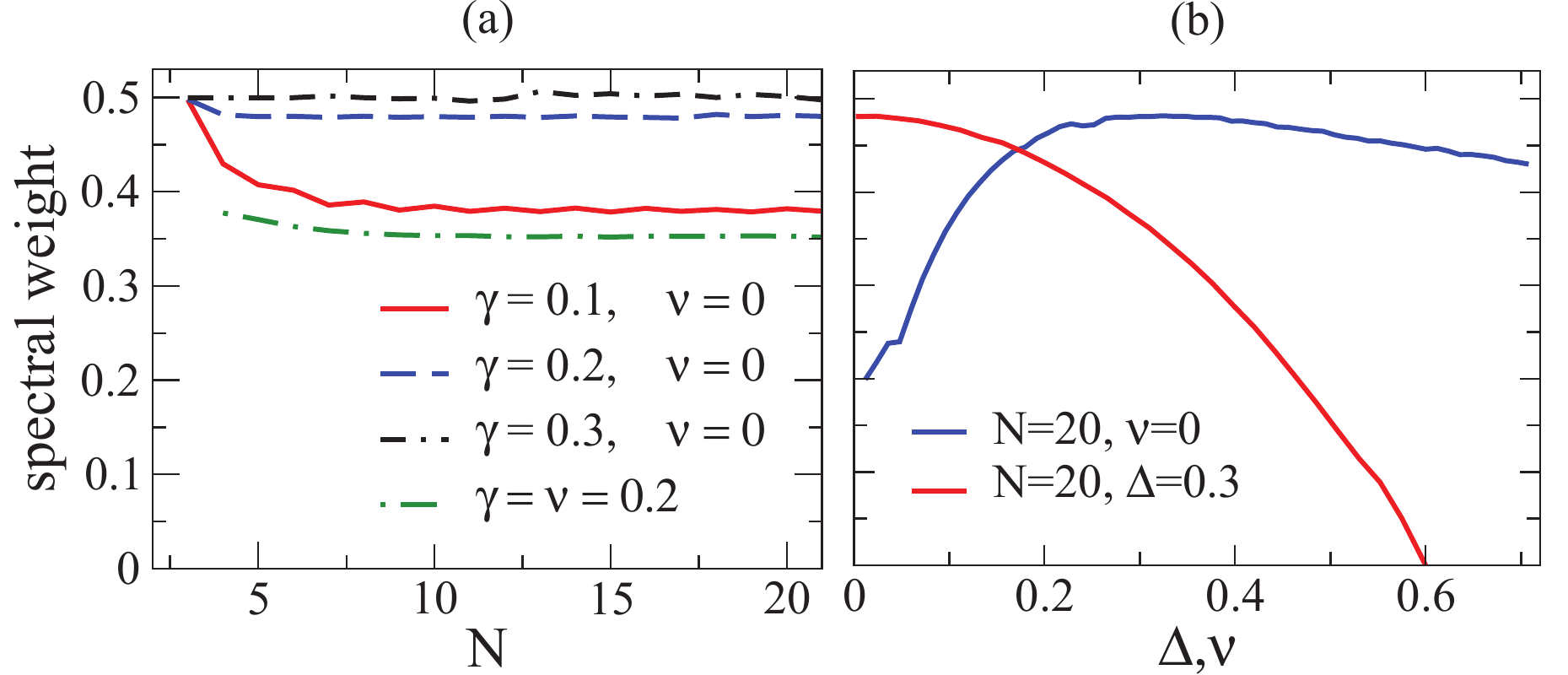}
\end{center}
\caption{Normalized spectral weight of the MBSs for $\gamma_{p}=10\,\gamma'_{p}=1$ ($p=L,R$), and (a) $\Delta=0.3$ or (b) $\gamma=0.3$. }
\label{sw1}
\end{figure}

We find that the normalized spectral weight of the MBSs initially decreases with increasing wire length [see Fig.~\ref{sw1}(a)] but quickly saturates to a length-independent finite value (cf.~with the ideas of electronic teleportation in these systems \cite{Semenoff07,*Fu10}) which depends on $\gamma$ and $\Delta$ but is independent of the strength of the contacts.
As can be seen in Fig.~\ref{sw1}(b), the maximum spectral weight for long wires is $50\%$ (a remarkably large fraction carried by a single pair of states) and is achieved when $\gamma=\Delta$ and $\nu=0$ when the two MBSs have exactly zero energy, are completely localized (one at each end) and decoupled from the rest of the wire (see Fig.~\ref{MajChain}). The contribution of the MBSs (which can be interpreted as the order parameter of the topological phase) decreases for increasing $\nu$ and at the topological QPT goes to zero in the way shown in Fig.~\ref{sw1}(b).

If we depart from the symmetric case considered above by setting $\mu_R=0$, we find that $j_{\mathrm L}$ gives an I-V similar to that of the symmetric case while $j_{\mathrm R}$ becomes zero (in agreement with Ref.~\onlinecite{Akhmerov11}).
We have also verified that moderate asymmetry between the left and right baths or the respective contacts does not significantly affect the results for the current as compared to the symmetric case and is thus meaningful to relate these to existing or future measurements.

In conclusion, by carrying out an exact microscopic calculation of non-equilibrium electrical transport in a \textit{p}-wave (Kitaev) chain connected to baths, we have been able to go beyond the previous works using either effective models of MBSs \cite{Bolech07,Nilsson08,Flensberg10} or a continuum representation such as the Bogoliubov-deGennes formalism \cite{Sengupta01}. First, we have captured the interplay and differences between Majorana and regular complex fermions (like the Bogoliubov excitations in the superconductor) and overcome the limitation of having voltages smaller than the SC gap. 
Second, we have modeled the contacts more realistically and with greater detail (going beyond the scope of scattering-formalism approaches; cf.~Refs.~\onlinecite{Nilsson08,Sengupta01}).
Third, we have shown how the microscopic properties and correspondingly the differential conductance varies with the length of the SC quantum wire for different parameter regimes. Fourth, we have compared and contrasted the behaviors when the number of sites in the wire is even \textit{vs.}~odd.
Fifth, we have calculated the normalized spectral-weight contribution of the MBSs to the total differential conductance and thereby provided a starting point to meaningfully compare the contributions to the transport coming from MBSs \textit{vs.}~other excitations in realistic experiments. And finally, we have predicted how the differential conductance changes across the topological QPT and shown the way in which the spectral-weight of the MBSs continuously goes to zero at the QPT. 

As shown in this letter, all these points not only make concrete experimental predictions, but also have strong implications for the scope and validity of the low-energy effective models.
Key features of the experimental $dI/dV$ curves \cite{Ha03,Sasaki11} are in
agreement with our results, in particular the overall shape of the zero-bias
conductance peak and the pronounced dips on its sides (which reflect the
plateaux shown in Fig.~\ref{I-V}). We suggest systematic measurements of the differential conductance based on our predictions and, in particular, the analysis of the spectral weight of the zero-bias conductance peak (such study would require a careful material-dependent protocol to take into account the effect of higher bands and dimensionality when defining the fractional spectral weight) as a way to strengthen and advance the experimental efforts to observe MBSs. 

\begin{acknowledgments}
We acknowledge the financial support by the University of Cincinnati. CJB \& NS are grateful for the hospitality of the Tata Institute of Fundamental Research (TIFR) where the last part of the research was done and a major part of the manuscript written.
\end{acknowledgments}

\bibliographystyle{apsrev-nourl}
\bibliography{Majorana}

\end{document}